\documentclass[twoside,12pt]{article}
\usepackage{epsfig}

\newfont{\sansb}{cmssbx10}
\newfont{\sans}{cmss10}

\begin{document}
{
\LARGE
High-Energy Interactions and Extensive Air Showers\footnote{Rapporteur talk
given at the 25$^{th}$ International Cosmic Ray Conference, Durban, South Africa, 1997}
\vspace{6pt}\\}

J. Knapp\footnote{E-Mail: knapp@ik1.fzk.de}
\vspace{6pt}\\
{\it 
Institut f\"ur Experimentelle Kernphysik, Universit\"at Karlsruhe, \\
P.O. Box 3640, D-76021 Karlsruhe, Germany
\vspace{-12pt}\\}
 
{\center ABSTRACT\\} 
In this report a summary of recent developments in the fields of high-energy
nuclear interactions (HE 1) and air shower phenomenology (HE 2) is presented.
New results from accelerator and cosmic-ray experiments and the progress in the
theoretical understanding and simulation are reviewed and their impact on air
shower analysis is discussed.

\setlength{\parindent}{1cm}
\section{Introduction}
The main astrophysical questions related to high-energy cosmic rays (CR) are those
for the source and the acceleration mechanisms of CR and the origin of
the knee in the all-particle spectrum. These questions can only be addressed if
the mass and the energy of the CR arriving on Earth can be determined.
Unfortunately, cosmic radiation at energies above 10$^{15}$ eV can
presently only be investigated by studying of extensive air showers (EAS) which
are produced by subsequent interactions of the primary particles with the nuclei
in the atmosphere.  Thus, the mass and the energy of the
primary particle are deduced from the measurable properties of the EAS.

The shower development depends on the primary mass and energy. However, it is
influenced by the properties of the hadronic and electromagnetic
interactions and by the mechanisms of the transport of secondary particles through the
atmosphere as well. Especially the hadronic and nuclear interactions impose large
uncertainties since they are only poorly known in the energy and kinematic
ranges of interest.  In addition, a detector with its limited acceptance and
efficiency gives a distorted picture of the secondary particles from which the
EAS properties have to be reconstructed.

The challenge of experimental EAS physics is to understand the shower
development and the detector performance well enough to enable a reliable
reconstruction of the mass and the energy of the primary particles starting from
the measured energy deposits and time signals in the single detector elements
without being able to calibrate the experiment in a suitable test beam with
well-known energy and mass composition.

In this article we try to highlight the recent developments in the understanding
of high-energy interactions, air shower development, and
reconstruction algorithms and discuss the implications on the interpretation of
experimental findings.
 

\section{New Results from Accelerator Experiments}
\subsection{Heavy Ion Collisions}
In the first collision of a CR particle with an air nucleus, 
a projectile nucleus (p, He, C,
... Fe) with energies of several TeV up to 10$^{20}$ eV hits a target nucleus
(N, O, Ar) with a random impact parameter.  Projectile and target are divided into
so-called participant nucleons, which are actively involved in the
collision, and spectator nucleons, which move on with their initial velocity.  
The participant nucleons contribute part or all of their
energy to form a fireball leading to the production of secondary particles,
while the spectators form remnant nuclei in more or less exited states that
consequently stabilize in a nuclear fragmentation process.  Most collisions are
peripheral with only few nucleons of projectile and target participating, few are
central with much energy involved in the particle production.

This kind of reactions is traditionally investigated in heavy ion collisions,
however, at much lower energies.  At CERN measurements have been performed with
Pb nuclei of $E= 158$ GeV/n and at Brookhaven with Au nuclei of 10.6 GeV/n
colliding with various targets.

New results on nuclear fragmentation \cite{HE1.1.2, HE1.1.3, HE1.1.9,
HE1.1.10} and on particle production \cite{HE1.1.6} have been reported. The
fragmentation of the projectile remnant is of special importance for CR physics
because in case of an air shower it carries a big part of the primary energy
through the atmosphere and, hence, influences the further shower development.
The KLM Collaboration, e.g., measured the emerging fragments by means of
lead-emulsion chambers with good spatial and charge resolution and presented the
distributions of the total charge of multicharged ($Z>2$) fragments in Au-H,
Au-(C,N,O), Au-(Ag,Br), and Pb-Pb interactions \cite{HE1.1.3}.  They found more
smaller fragments for more symmetric projectile-target systems and for heavier
targets.  In 10.6 GeV/n Au-emulsion collisions \cite{HE1.1.9} a
characteristic dependence of the distribution of the total charge bound in
fragments ($\sum Z$ with $Z>2$) on the average transverse momentum
$\langle p_\perp\rangle$ is observed.  
Events with $\langle p_\perp\rangle > 90$ MeV/c
coming from more central collisions exhibit basically a flat distribution of
$\sum Z$ while events with $\langle p_\perp\rangle < 90$ MeV/c from peripheral
interactions show a pronounced peak at $\sum Z = 65 ... 80$ because in peripheral
collisions a big projectile survives.  Those distributions, depending on energy
and on the mass of projectile and target, are important to check and tune the
models being employed for relativistic nuclear reactions.

The production of secondary particles in the central collision zone is
investigated in Ref. \cite{HE1.1.6}. Their distribution in
pseudorapidity $\eta = \frac{1}{2} \ln \frac{|p|+p_{_L}}{|p|-p_{_L}} =
- \ln \tan (\theta/2)$, which basically represents the emission angle of the
particle to the beam,
is compared to calculations with the FRITIOF 
\cite{fritiof4,fritiof} and the VENUS \cite{venus}
model. The latter model allows for interactions of secondaries particles with
spectators or with each other in case of large multiplicities leading to a
redistribution of particle energies and a better agreement with experimental
data at lower energies and with lower projectile and target masses
\cite{venus}. Re-interactions become more important for higher energy
densities, i.e. larger nuclei and more central collisions.  In Pb-Pb collisions,
however, it turns out that all models overestimate the central
pseudorapidity density by 30 to 50\% and VENUS without re-interaction agrees
better with the data than VENUS including it.  These findings indicate that the
description of particle production in the models still is to be improved for
heavy nuclei.
 
\subsection{HERA Results}
Events with high $p_\perp$ secondaries or jets become
more important
with increasing energy.  
While low $p_\perp$ secondaries still dominate by far, the tail
of high $p_\perp$ particles is growing with energy, giving rise to the so-called
minijets with $p_\perp$ in the few GeV range.  The rise of the cross-section for
minijet production is strongly dependent on the structure of the nucleon, since
with increasing energy more and more low-energy partons of projectile and target can
contribute to inelastic collisions. 
Gluons dominate due to their vanishing restmass at low
parton energies.
The distribution of parton momenta $x =
p_{\rm parton}/p_{\rm total}$ inside a nucleon is described by the structure
function $F_2$ which was measured recently at the HERA $e$-$p$ collider for
different momentum transfers $Q^2$ down to $x \approx 10^{-5}$ (see
e.g. Ref. \cite{h1pap}). 
Fig. 2 shows that the experimental results from the H1
experiment are in good agreement with parameter-free calculations from first QCD
principles by Gl\"uck, Reya, and Vogt \cite{grv} and  several other calculations.
These structure functions for the first time allow to calculate the minijet
cross-sections up to highest cosmic-ray energies on a reliable theoretical basis
and should be implemented in all high-energy hadronic interaction generators.

\subsection{p-$\bar{\rm \bf p}$ Collider Experiments}
\label{sec-minimax}
The MINIMAX experiment \cite{HE1.1.15} at the Fermilab collider was designed
to measure secondaries in very forward direction at a pseudorapidity of $\eta =
4.1 \pm 0.3$.  Its primary physical goal is the search for disoriented chiral
condensates (DCC) \cite{bjorken} which manifest themselves by large fluctuations
in the fraction $f = N_{\pi^0}/(N_{\pi^0}+N_{\pi^\pm})$ of $\pi^0$s produced in
an interaction. For DCC the distribution of $f$ is proportional to
$0.5/\sqrt{f}$ and not binomial as predicted by standard interaction models (see
Fig. 3).
Therefore, DCC have been discussed as a possible source of Centauro-type events.  
These are events with a large number of hadrons
and a comparatively small number of photons. Centauro-type events have been detected
by several emulsion chamber experiments and are interpreted as an indication of new
physics in hadronic interactions.

The MINIMAX Collaboration has analyzed $1.5\times 10^6$ events at $E_{\rm cm} = 1.8$
TeV corresponding to a lab energy of 2 PeV.  A set of observables $r_i$
with
$$ r_i = \frac{\langle 1-f\rangle\langle f(1-f)^i\rangle}{\langle
f\rangle\langle (1-f)^{i+1}\rangle} $$ 
has been defined which are rather insensitive to experimental errors. 
They all are $\equiv 1$ for binomial fluctuations and
$1/(i+1)$ for pure DCC events.  Tab. \ref{tab-dcc} shows the result in
comparison with PYTHIA calculations \cite{pythia} including 5\% and 10\% DCC events.
This comparison shows no evidence for DCC in the data, i.e. at the level $\le 1\%$.  
The results rather suggest $r_i \ge 1$ \cite{minimax}.  

Since this result is obtained from $p$-$\bar{p}$ collisions, it does not exclude
large fluctuations in $f$ for nucleus-induced collisions. A recent analysis of
158 GeV/n Pb-Pb collisions by the WA98 Collaboration, however, sets 
only upper limits of the DCC content in their data \cite{wa98}.

\begin{table}[h]
\caption{\small Results from the MINIMAX experiment
compared with Monte Carlo calculations including 5\% and 10\% DCC contribution and
with expectations for pure DCC.}

\label{tab-dcc}
\begin{center}
\begin{tabular}{|c|c|cc|c|}
\hline 
 Obs.   & Data & \multicolumn{2}{|c|}{Monte Carlo} & DCC \\
        &      &  5\% DCC & 10\% DCC &  \\
\hline
 $r_1$ & 1.0228$\pm$0.0035 & 0.97$\pm$0.02 & 0.95$\pm$0.02 & 0.5   \\
 $r_2$ & 1.0320$\pm$0.0097 & 0.93$\pm$0.05 & 0.89$\pm$0.04 & 0.333 \\
 $r_3$ & 1.0563$\pm$0.0251 & 0.95$\pm$0.10 & 0.89$\pm$0.08 & 0.25  \\
\hline 
\end{tabular}
\end{center}
\end{table}

Further analyses of MINIMAX data are in progress.  Soon, better limits on
DCC, pseudorapidity distributions for charged and neutral pions, and data on
$K^0$ and $\Lambda^0$ production will be available.

\vspace*{-3mm}
\subsection{Future Accelerator Experiments}
With the FELIX experiment a full acceptance detector is proposed at the Large
Hadron Collider LHC \cite{HE1.1.14}.  It aims for full coverage in $\eta$ and
$\phi$ for charged and neutral particles and will reuse the magnets of ALEPH,
UA1, D0, RHIC DX, and UNK in different rapidity ranges. A schematic view is
shown in Fig. 4.  With this set of magnets the experiment will
have a reasonably uniform $p$ and $p_\perp$ resolution over the whole acceptance.
FELIX will be able to investigate particle production beyond $\eta = 7$. It is
well suited for small angle physics, hard and soft QCD phenomena in $p$-$p$,
$p$-$A$ and $A$-$A$ collisions, and for search for cosmic-ray anomalies.

With the start of operation of the Relativistic Heavy Ion Collider (RHIC) at
Brookhaven in spring 1999, another facility will become operational that can
deliver new insights into the physics of heavy ion collisions.  Experiments will
search especially for signatures of the onset of new physics in central Au
collisions at 200 GeV/n, e.g. the phase transition to quark gluon plasma
\cite{HE1.1.16}.

\section{Interaction and EAS Models}
Presently there are several Monte Carlo programs being used for the full
simulation of EAS development in the atmosphere. The most common ones and the
models used for high-energy hadronic interaction are shortly described in the
following sections.

\subsection{Available Codes}
{\bf CORSIKA} \cite{corsika_phys,corsika_users,HE2.3.2} is a multi-purpose
shower simulation program of air shower development. It contains the hadronic
interaction models VENUS \cite{venus}, QGSJET \cite{qgsjet}, DPMJET
\cite{dpmjet}, SIBYLL \cite{sibyll,superpos}, and HDPM
\cite{capdpm,corsika_phys} at high energies and GHEISHA \cite{gheisha} and EGS4
\cite{egs4} for low-energy hadronic and electromagnetic interactions,
respectively.  VENUS, QGSJET, and DPMJET base on the Gribov-Regge theory of 
multi-Pomeron exchange, which has been used successfully over decades to describe
elastic and inelastic scattering of hadrons.  Especially nucleus-nucleus
collisions and diffraction are treated in great detail in these models.  SIBYLL
is a minijet model that describes the rise of the cross-section with energy by
increasing the pairwise minijet production.  HDPM uses parametrizations along
the collider data for $p\bar{p}$ collisions. The latter two models apply the
Glauber theory for hadron-nucleus collisions and treat projectile nuclei as a
superposition of free nucleons.  With the appropriate interaction models CORSIKA
can simulate air showers up to $10^{21}$ eV. A thinning algorithm according to
Hillas keeps computing time and disk space to a manageable level.
CORSIKA versions exist for Cherenkov light production, horizontal showers, neutrino
generation and several other special purposes.  All parts of CORSIKA are
described in detail and are available to the cosmic-ray community.

\noindent
{\bf HEMAS} \cite{hemas,hemas2} was originally developed for the MACRO and
NUSEX experiments on the basis of the SHOWERSIM program \cite{showersim} with
extensions to energies of around $10^{17}$ eV. These experiments only register
high-energy muons underground and, therefore, only the simulation of the 
high-energy part of the air shower and the lepton production had been emphasized.
Detailed transport of TeV muons through rock \cite{hemasrock} was implemented
and nuclear reactions were treated under the superposition assumption.  In
the most recent version of HEMAS the hadronic event generator was replaced by
DPMJET including an elaborate algorithm for nuclear fragmentation \cite{fzic}.
In addition, the low-energy part of the cascade development was improved for the
use with the EAS-TOP experiment measuring the low-energy shower particles.

\noindent
{\bf MOCCA} \cite{mocca,mocca1} is one of the first EAS programs and was
developed by M. Hillas in the 1980s. Its original hadronic
interaction model is the so-called splitting algorithm which produces
secondaries following a very simple qualitative prescription with 2 free
parameters. To overcome the limitations of this simple event generator, the more
elaborate SIBYLL code has been implemented. MOCCA can only deal with protons,
neutrons, pions, kaons, electrons, and photons, neglecting all excited and 
short-lived states that can be produced in high-energy interactions.  Due to a
statistical thinning algorithm, MOCCA is able to simulate showers up to highest
energies. Mean values are reproduced well, but the fluctuations of shower
quantities are grossly overestimated when thinning too much.  MOCCA optionally
performs a simple simulation of the reactions of the shower particles with the
detector materials.  Using other EAS programs this part of the simulation has to
be done in a separate step, e.g. with the well accepted detector Monte Carlo
program GEANT \cite{geant}.

\noindent
{\bf AIRES} \cite{aires} presently is a translation of MOCCA into standard
Fortran with much of the code being restructured and documented. The detector
simulation part has been omitted. It is envisaged for the future to extend the
program by other interaction models and unstable particles, and to improve the
algorithms for particle transport, decays and shower development.

There are some more programs in use in the air shower community.  But these are
either special purpose programs or private versions of one experiment, 
the physics input of which is not published in detail and which are not generally
available.

\noindent
{\bf GENAS} \cite{genas} is constructed for high-speed calculation of EAS in the
10$^{20}$ eV range.  Electromagnetic subcascades are computed by the {\bf COSMOS}
program \cite{cosmos}.  GENAS/COSMOS is used to interpret the AGASA data.
It is not a full Monte Carlo program, but uses parametrizations for electron and
photon numbers as a function of core distance for different atmospheric
depths. 
Only 6 groups of 
projectiles are distinguished, from proton \& He, light
heavy ($\langle A\rangle =8$), medium heavy ($\langle A\rangle =14$), heavy
($\langle A\rangle =25$), very heavy ($\langle A\rangle =35$) to
iron. Secondary muons and hadrons are neglected.

\noindent
{\bf MC0} and {\bf MJE} are used for the interpretation of results obtained by
the emulsion chamber experiments at Pamir (see
e.g. Refs. \cite{HE1.2.15,HE1.2.18,HE1.2.21}). They are based on the Quark Gluon
String model with minijet production. However, their capabilities and
limitations are not well described in literature.

Hadronic interaction models exist which are used for high-energy
accelerator experiments.  PYTHIA \cite{pythia,sjoestrand} models hadronic
interactions with high momentum transfer according to QCD, takes into account
resonance formation as well as gluon radiation from quarks and contains the
fragmentation of colour strings into colour neutral hadrons.  It also contains 
the soft (minimum bias) processes, which are important for air showers, but
cannot handle primary mesons or nucleus-nucleus collisions.  FRITIOF
\cite{fritiof4,fritiof} 
is able to simulate hadron-nucleus and nucleus-nucleus
collisions on the basis of the classical string theory.  The results for hadron
interactions are similar to those of Gribov-Regge models.  For the treatment of
nucleus-nucleus collisions FRITIOF adopts the superposition model.  A
combination of FRITIOF and PYTHIA could, in principle, be used to simulate EAS.

\subsection{Model Comparisons}
The major systematic uncertainties in EAS analysis arise from the lack of
knowledge of the total cross-sections and the details of particle production
for nuclear and hadronic reactions at high energies with small momentum
transfer.  The experimental findings at proton colliders have to be extrapolated
over many orders of magnitude to CR energies, small emission angles and nuclear
projectiles and targets.

To estimate the systematic errors, a detailed comparison of hadronic interaction
models inside the frame of the CORSIKA program has been performed \cite{HE1.3.3}.
The models involved were VENUS, QGSJET, DPMJET, SIBYLL and HDPM.  Their basic
properties are summarized in Tab. \ref{tab-mods}.
\begin{table}[b]
\caption{\small Basic features of the interaction models used in CORSIKA.}
\label{tab-mods}
\begin{center}
\begin{tabular}{|l|ccccc|}
\hline 
                   &  HDPM   &  SIBYLL &  QGSJET &  VENUS  & DPMJET \\
\hline
Gribov-Regge       &         &         &    +    &    +    &   +    \\
Minijets           &         &    +    &    +    &         &   +    \\
Sec. Interactions  &         &         &         &    +    &        \\
N-N Interaction    &         &         &    +    &    +    &   +    \\
Superposition      &   +     &    +    &         &         &        \\
Max. Energy (GeV)  & 10$^8$  &$>10^{11}$&$>10^{11}$& 5$\times$10$^7$& 10$^9$ \\
Memory (Mbyte)     &  7.6    &   8.3   &   9.7   &   21    &   52   \\
CPU Time$^1$ (min) & 500     &  371    &  484    &  2250   & 1310   \\
\hline 
\multicolumn{6}{l}{\footnotesize $^1$ 
500 p showers at 10$^{15}$ eV,
$E_{\mu},E_{h} > 0.3$ GeV, 
110 m a.s.l., NKG option, on DEC-AXP 3600 (175 MHz)} \\
\end{tabular}
\end{center}
\end{table}

\subsubsection{Cross-Sections}
The first quantities compared are the inelastic p-air cross-sections
$\sigma_{\rm inel}^{\rm p-air}$.  All models except for SIBYLL calculate
$\sigma_{\rm inel}^{\rm p-air}$ from the experimentally well determined
$\sigma_{\rm inel}^{\rm \bar{p}-p}$ by assuming a distribution of the nucleons
in the air nuclei.  Therefore, all these models agree reasonably well with each
other (and the collider data) and start to diverge only at energies where no
measurements exist anymore.  The calculation of $\sigma_{\rm inel}^{\rm p-air}$,
however, leads to rather different results as shown in
Fig. 5, where the cross-sections are plotted together with
experimental data from cosmic-ray experiments.  HDPM predicts the smallest
cross-section exhibiting the flattest rise with energy.  The Gribov-Regge type
models show a comparable rise but differ by about 50 to 100 mb. Generally, they
are below the experimental points at high energies.

SIBYLL has adopted a parametrization for $\sigma_{\rm inel}^{\rm \bar{p}-p}$
which at $p_{\rm lab} < 10^6$ GeV/c exhibits the flattest rise with values
clearly below the experimental results.  At $p_{\rm lab} > 10^6$ GeV/c it rises
steeply and surpasses all other models.  As a consequence, $\sigma_{\rm
inel}^{\rm p-air}$ in SIBYLL also rises more steeply than in any other model, fitting
best the data points in Fig. 5.  The MOCCA cross-section
grows even faster than the SIBYLL one with up to 100 mb higher values at
$p_{\rm lab} \approx 10^8$ GeV/c.

In the most recent version of DPMJET (II.4, not yet available in CORSIKA)
$\sigma_{\rm inel}^{\rm p-air}$ has been corrected downwards by 25 mband now
agrees nicely with the cross-sections of VENUS and QGSJET at low energies.  A
new analysis of the EAS-TOP Collaboration determined the p-air cross-section at
$p_{\rm lab} \approx 2\times 10^6$ GeV/c to be $\sigma_{\rm inel}^{\rm p-air} =
344 \pm 12$ mb \cite{HE1.2.1}, in contrast to the high values of
Ref. \cite{gaisser_cross} and in reasonable agreement with the predictions of
the Gribov-Regge models.

The spread between the models amounts to about 100 mb or 25\% in the region of
the knee of the all-particle spectrum.  Since the inelastic cross-section
determines the mean free path of a particle in the atmosphere, it influences
directly the longitudinal shower development. A larger cross-section causes
shorter showers and, consequently, fewer particles at ground level.  The
differences in $\sigma_{\rm inel}^{\rm p-air}$ for comparable assumptions of
$\sigma_{\rm inel}^{\rm \bar{p}-p}$ originate partially from different
applications of the Glauber theory and from varying assumptions regarding the
form of the target nuclei.  The discrepancies between the models are rather big,
taking into consideration that all authors use basically the same approach to
calculate the cross-sections.  By agreement on the {\em best} method of
calculation, a big part of the discrepancies should vanish.

\subsubsection{Particle Production}
The production of secondaries in hadronic interactions also differs between
models. A variety of quantities has been examined, the results are described
in Refs. \cite{modcomp,HE1.3.3}.  The quantity with the largest impact on
air shower development is the inelasticity, i.e. the fraction of the energy of a
projectile that is used for production of secondary particles.  Again, a
variation in this quantity directly implies a modification of the longitudinal
shower development.  The inelasticity does not vary much with energy, but
differences between the models amount to 20 to 30\%.

The effects of inelasticity and cross-sections are basically independent and may
cancel out or add up. For DPMJET with the largest inelasticity and the largest
cross-sections the showers are very short, while they are longest for HDPM with
the smallest inelasticity and the smallest cross-sections. This leads to
differences between the predictions of the electron or muon number in a $10^{15}$ eV
EAS at sea level of about a factor of 2, or, conversely, to a corresponding 
systematic error
in the shower energy when it is determined from the particle
number at ground level only \cite{HE1.3.3}.

Similar comparisons of the MOCCA and SIBYLL interaction generator inside AIRES
and of DPMJET and the older interaction model inside HEMAS are in preparation and
will soon shed additional light on the systematics of hadronic interaction models.

\subsubsection{Other Comparisons}
Besides a comparison of different interaction models in the same program frame,
comparisons of different frames with identical hadronic interaction models are
of interest.  They reveal the systematics of the electromagnetic
interactions, particle transport, decays, treatment of low-energy particles and
the whole philosophy of the simulation.  Such tests are in progress by comparing
CORSIKA with SIBYLL and MOCCA or AIRES with SIBYLL, CORSIKA with DPMJET and
HEMAS with DPMJET, and CORSIKA with QGSJET and MC0.
The CORSIKA/SIBYLL showers, e.g., develop about
10-20 g/cm$^2$ higher in the atmosphere than the corresponding MOCCA/SIBYLL
showers.

\subsection{Improvements}
The model comparisons so far have brought the systematics of EAS analysis to the
attention of the community and to the authors of the interaction models. It has
triggered already a series of improvements.
There is a new DPMJET (version II.4) available with modified cross-sections and an
improved treatment of nuclear fragmentation.  The authors of SIBYLL began
to revise their code and the cross-sections used.  The authors of VENUS
and QGSJET started a collaboration to unify their codes by combining the best
parts of each program and forming a model to the best of the present knowledge
\cite{venusneu}. The new VENUS/QGSJET should be available by the end of 1997.
The HDPM model obtained an improved $p_\perp$ generator taking into account the
correlations of $\langle p_\perp\rangle$ and the central rapidity density and a
better treatment of nuclear fragmentation by means of an abrasion/evaporation
algorithm \cite{HE1.3.1,HE2.3.23}.

\subsection{Further Simulation Activities}
A series of other new activities around the shower simulation were reported
recently.

Gaisser, Lipari and Stanev \cite{HE2.3.15} presented a new, fast, 1-dim. shower
calculation to obtain the longitudinal profiles of 10$^{20}$ eV showers by using
precalculated lower-energy pion-induced showers and a bootstrap method. It
allows to examine quickly the effects of details of the interaction model as has
been demonstrated in Fig. 1 of Ref. \cite{HE2.3.15}.

Vassiliev et al. \cite{HE2.3.5} presented a new algorithm for a faster and
better calculation of the Cherenkov light production from electromagnetic
subshowers of EAS. The algorithm, being specialized for only one energy range of
interest, uses extensively lookup tables and a thinning technique for the
Cherenkov photons.

Konopelko and Plyasheshnikov \cite{kono} advocate a semianalytical Monte Carlo
technique (SAMC) for speeding up high-energy shower calculations.  In contrast
to most other approaches they solve the adjoint cascade equations for the 
high-energy part of the shower analytically and perform a full Monte Carlo simulation
for the low energies. The authors claim that
with this approach the fluctuations of shower quantities
can be accounted for properly while this is not the case for other thinning
techniques or hybrid Monte Carlo programs.

Capdevielle et al. \cite{HE2.3.23} elaborated a fast method for simulation of
the electromagnetic part and the fluorescence light of highest-energy air
showers.

\subsection{Impact on EAS Analyses}
The impact of the above-quoted systematic discrepancies between different EAS models
can be demonstrated for the results on CR composition in the energy
range from $10^{17}$ to $10^{19}$ eV obtained by the Fly's Eye and the Akeno
experiments.

The Fly's Eye detector consists of many photomultipliers, each observing a small
segment of the night sky. The charged particles in a shower induce nitrogen
fluorescence in the air which can be registered over large distances for primary
energies above $10^{17}$ eV as a track across several photomultiplier pixels.
From the signals in each photomultiplier the light curve is reconstructed.
Since the fluorescence light is mainly produced by the large number of low-energy
electrons, the light curve traces closely the energy deposition in the
atmosphere, and, hence, the longitudinal shower development in a calorimetric
way.  The integral over the light curve is a good measure of primary energy
and the height of the 
maximum of the shower development relates to the mass of the primary
particles. In
1993, the Fly's Eye Collaboration published the average depth of shower maximum
as a function of energy and found a change with energy.  
The elongation rate $dX_{\rm max}/dE$ was larger than the predictions by their KNP
model for pure compositions of either protons or iron nuclei
\cite{flyseye_comp,gaisser_d47}.  On the basis of these simulations it was
concluded that the composition changes from almost pure Fe at $E \approx
10^{17}$ eV to nearly pure p at $E \approx 10^{19}$ eV.

However, the other experiment which observes EAS in this energy range
could not confirm the Fly's Eye result.
The AGASA experiment at Akeno is a 100 km$^2$ scintillator array with 111
electron detectors and 27 muon detectors for $E_\mu > 0.5$ GeV.  The
density of charged particles $\rho_{600}$ at a core distance of 600 m was used
as energy estimator and the muon density at the same distance served
as mass indicator.  In the AGASA data a change of composition should have led
to a change in the ratio of the observed muon density to the one expected for
constant composition.  No such change was observed.  The analysis, however, was
based on calculations with the GENAS/COSMOS model.  This claim was affirmed by a
new analysis \cite{HE2.2.10} of a dataset with 7.5 times the statistics of the
old experiment. The mass change is $< \pm 5$ in the energy range from $E = 3
\times 10^{17}$ to $10^{19}$ eV. Their result stays basically unchanged when
employing MOCCA/SIBYLL for the analysis \cite{HE2.2.10}.

Are the Fly's Eye and the AGASA results really contradicting each other?
It was soon emphasized that the different conclusions just reflect the
differences in the Monte Carlo programs used.

Kalmykov et al. \cite{kalm5} compared the Fly's Eye data with data from the
Yakutsk array \cite{yakutsk} and with calculations using their QGSJET program
(see Fig. 6).  They found good agreement between the experiments
and with the simulations when assuming the ($\sum$) composition of light :
medium : heavy $\approx$ 0.55 : 0.22 : 0.23 at $E < 3\times 10^{15}$ eV with a
very smooth change towards heavier nuclei leading to 0.23 : 0.22 : 0.55 at $E =
10^{17}$ eV.

Gaisser et al. \cite{HE2.3.15} showed that SIBYLL gives a somewhat larger
elongation rate than the KNP model and, therefore, leads to a modified conclusion 
concerning the change of mass composition.

On the other hand, Dawson et al. \cite{OG6.1.10} compared data of the A1 stage
of the AGASA array with MOCCA/SIBYLL simulations and found a similar change of
composition as described in the original Fly's Eye publication (see
Fig. 7).  The calculations were performed for a muon energy
threshold of 1 GeV, corresponding to the threshold of the muon detectors.  In
contrast, the analysis described in \cite{HE2.2.10} used simulations with a muon
threshold of 0.5 GeV and corrected for the mismatch. This
may be the reason for the apparent discrepancy of the results of
Refs. \cite{HE2.2.10} and \cite{OG6.1.10}.

As a conclusion, the situation in EAS physics is presently like the one sketched
in Fig. 8 : better use the same {\em yardstick} (i.e. Monte
Carlo program) to get {\em consistent} results in different experiments, and 
use a well calibrated, reliable yardstick to obtain the {\em correct}
result.  In other words, the interpretation of EAS data strongly depends on the
air shower model used, as long as the models differ the way they presently do.

For progress in the interpretation of EAS measurements, we feel that it is
vitally important to make a common effort towards a reference simulation program
that contains the best and most detailed treatment of all physical processes
relevant to EAS and that is used and tested by groups working in all parts of
cosmic-ray physics without adapting the physics parameters for each experiment
in a different way.  There is a very successful example of such an activity. The
CERN detector simulation package GEANT \cite{geant} has evolved over the past
decade to one of the most powerful and efficient tools of high-energy physics
experiments.  By the contribution of many users and experiments, the algorithms
for each aspect of the simulation have 
been refined continuously and have
acquired a quality which is beyond the reach of a single person or a small
group. It is nowadays the de facto standard for the simulation of the detailed
response of detectors in complex setups to the incidence of all kinds of
radiation or relativistic particles.

Such a reference should also serve to estimate the performance of special
purpose programs that are optimized for particular aspects of CR physics, such
as fast calculations of highest energy showers, TeV muons, Cherenkov light
production, and so on.

\section{Air Shower Arrays}
\subsection{Multiparameter Measurements}
Air shower analyses are based on the comparison of experimental data with MC
simulations. To be able to perform such a comparison, a spectral form, an
energy-dependent mass composition and parameters of the high-energy interactions
have to be assumed. Therefore, a discrepancy between MC and data can have many
sources and, on the other hand, an agreement does not necessarily mean that {\em
all} the assumptions are right.  Especially when registrating only one
observable (e.g. $N_e$) several parameter settings may exist that can reproduce
the observation. Fluctuations in the observable are then directly projected onto
uncertainties in primary energy or mass.

When measuring several quantities it is possible to recognize fluctuations.
A big part of the shower fluctuations originates from the first hadronic
interaction. If, by chance, in this interaction more than average $\pi^0$s are
produced, more energy is transferred to the electromagnetic and less to the
hadronic and muonic component.  An energy estimate based on the electromagnetic
energy only, therefore, will overestimate the primary energy.  Taking all
components into account, one realizes that the deficiency in one component is
partially compensated by an excess in another component and, consequently,
obtains a better energy estimate.  Similar arguments hold for the varying height
of the first interaction.

With multiple observables in addition correlations between shower variables
can be investigated which allow to test model assumptions more stringently and
to disentangle influences of different sources on the observables.

In recent years, the need of multiparameter measurements was recognized and many
of the air shower experiments were upgraded or designed to be able to measure
simultaneously as many shower variables as possible. The HEGRA experiment has
been extended with the AIROBICC Cherenkov counter array and with muon
detectors. CASA-MIA was completed by the DICE Cherenkov telescopes and the BLANCA
Cherenkov light detectors. The EAS-TOP Collaboration enhances the shower
information by the registration of high-energy muons seen in the underground
detectors in the Gran Sasso tunnel. KASCADE measures the electromagnetic,
muonic, and hadronic component with high resolution and high dynamic range.

Many results on spectra and composition of cosmic rays have been
presented at this conference, most of them, however, in the session OG
6. Therefore, they are summarized in the rapporteur article by A.A. Watson
\cite{watson} in this volume.

\subsection{EAS-TOP}
The EAS-TOP experiment at the Gran Sasso in Italy measures air showers at 2000 m
a.s.l. with a 10$^5$ m$^2$ scintillator array (35 stations, 10 m$^2$ each) and a
144 m$^2$ muon/hadron detector, consisting of 9 layers of tracking and
proportional chambers in an absorber of 6 $\lambda_I$ thickness.  With the
EAS-TOP detectors the electron number and the number of muons with $E_\mu > 1$
GeV can be determined.  Below the array the underground
experiments MACRO and LVD are run and coincidences with the array are
registered.  The rock shielding is equivalent to 3400 m of water absorbing the
electromagnetic, hadronic and most of the muonic part of EAS. Only muons with
energies exceeding 1.3 TeV can reach the underground detectors. Coincident
events allow to investigate additional parameters of an air shower in
correlation with the ones measured by the array and the muon/hadron detector.
Results of coincident measurements of EAS-TOP and LVD were reported in
Ref. \cite{HE2.1.8}.  LVD is a liquid-scintillator detector with 78 m$^2$ area
for a good dE/dx measurement combined with limited streamer tubes for good
tracking. Based on MC calculations, 
very stringent cuts in $N_e$ (measured by EAS-TOP)
and the high-energy muon number $N_{\mu~\rm TeV}$ (measured by LVD) were defined
to select p and Fe-enriched samples with a low efficiency but a high purity.
For the resulting event samples the number of low-energy muons as measured in
the array $N_{\mu~\rm GeV}$ were investigated.  The results in this independent
variable agree nicely with the Monte Carlo expectations for p and Fe-induced
showers as shown in Fig. 9.
This is a test of the EAS models involved and the agreement indicates a
good description for the observables used and their correlation.  Such
multiparameter measurements allow consistency checks and stringent tests of the
EAS models which are not possible with only one measured quantity.

As already mentioned a new measurement of the inelastic p-air cross-section was
presented by the EAS-TOP Collaboration \cite{HE1.2.1}.  It was evaluated by the
analysis of the rate of showers with the same electron and muon number for
different zenith angles and amounts to $\sigma_{\rm inel}^{\rm p-air} = 344 \pm
12$ mb at $p_{\rm lab} \approx 2\times 10^6$ GeV/c \cite{HE1.2.1}.  This value
is about 100 mb or 25\% lower than previous measurements
\cite{honda_cross,baltru_cross,gaisser_cross}.  What could be the reason for
this large discrepancy?  While the raw data of AKENO and EAS-TOP do not differ
much, the disagreement appears after applying a correction factor which accounts
for the bias in the attenuation length due to the analysis procedure.  In case
of EAS-TOP the correction is energy-dependent from 1.1 to 1.4. It was obtained
by a full MC simulation of the shower development, the detector response and the
reconstruction procedure. The AKENO Collaboration applies a constant correction
of 1.5. The difference in the correction factor has about the size of the
discrepancy in the cross-sections. This comparison suggests a strong dependence
of the result on details of the analysis procedure. The difference of 25\%
between EAS-TOP and AKENO may serve as an estimation of the systematic error.

With its muon/hadron detector the EAS-TOP Collaboration has measured the flux 
of single hadrons at 2200 m a.s.l. and found
$$ I(E_h) = (1.89 \pm 0.21) \times 10^7 \cdot (E/{\rm TeV})^{-(2.67\pm 0.14)}
\quad \mbox{/(m}^2\mbox{ s sr GeV)} \quad .$$ The data are shown in
Fig. 10. The flux of TeV hadrons is about 5 times higher than
the one measured near sea level \cite{mielke}.  From the comparison of the two
results an attenuation length of $\Lambda = 115.9$ g/cm$^2$ is deduced.
Under the assumption that a single hadron at ground level is a particle closely
related to a primary proton and survived the interactions as leading particle
the attenuation length $\Lambda$ is directly related to the interaction length
$\lambda_{\rm int}$ and the inelasticity $\eta$ of the interaction by the form
$\Lambda = \lambda_{\rm int}/(1-(1-\eta^\gamma))$ \cite{bellandi}.  $\gamma$ is
the slope of the primary energy spectrum.  Using this relation, the inelastic
proton air cross-section of $\sigma_{\rm inel}^{\rm p-air} \approx 290$ mb 
at
$E=1$ TeV, and the above-quoted value of $\Lambda$, 
an inelasticity of $\eta \approx
0.47$ is obtained.  
The quoted error of this estimate is $\pm 0.07$, but
the systematic uncertainties of the assumptions and the formula used are
probably larger than that. A MC study could reveal how well these simple
relations hold in a realistic environment with fluctuations and imperfect
detection systems.

\subsection{KASCADE}
Measuring different components in an EAS bears the problem of each detector 
designed for one shower component being sensitive to the others as well, to some
extent.  If the components are to be identified separately, e.g.  in a setup
with $e/\gamma$ detectors above and $\mu$ detectors below an absorber plate, the
energy depositions in the $e/\gamma$ detectors need to be corrected for
penetrating muons and hadrons and the signals of the $\mu$ detectors for the
electromagnetic and hadronic punch-through.  The correction, however, depends in
general on the type and energy of the primary particle as well as on the zenith
angle and the distance of a detector to the shower axis. The contributions of
the various components to the energy deposit were investigated for the array
electron and muon detectors of the KASCADE experiment \cite{HE2.1.4}.  A set of
appropriate correction functions was evaluated in an iterative way (see Figs. 1
\& 2 in Ref. \cite{HE2.1.4}) and finally 
meaningful values for the electron and muon numbers $N_e$ and $N_\mu$
were obtained. Several
analyses on the composition and energy spectra rely on 
$N_e$ and $N_\mu$ determined this way.

The KASCADE experiment finds a pronounced knee in the electron and the muon size
spectra \cite{HE2.1.5}.  The slopes of the electron size spectrum are $-2.45 \pm
0.01 \pm 0.05$ below and $-2.94 \pm 0.02 \pm 0.1$ above the knee and for the
muon size spectrum the corresponding numbers are about -3.0 and -3.4.  The
position of the knee varies with zenith angle \cite{HE2.1.5} in contradiction to
findings of the Tien Shan Collaboration \cite{HE2.1.22} which sees the knee for
all zenith angles at the same $N_e$. The KASCADE findings
are in agreement with the EAS-TOP results as
presented at the last European Cosmic Ray Conference \cite{eastop_perp}. With
improved statistics a two-dimensional analysis will become feasible. 

The unique part of KASCADE is a large hadron calorimeter. Measuring the hadron
component in addition to the muonic and electromagnetic components allows
multiparameter analyses with the use of correlations between single variables
and a sensitive test of the EAS models.

It has been noted earlier \cite{modcomp} that the SIBYLL model predicts fewer
muons for high-energy air showers as compared to other models. This leads to a
strong discrepancy between data and simulation, e.g. in the energy spectra of
hadrons as shown in Fig. 11 \cite{HE1.2.27}.  While the data
points lie between VENUS predictions for p and Fe-induced showers, leading to a
composition estimate between p and Fe, they lie outside the range of SIBYLL
predictions. Thus, with SIBYLL one would predict a composition much heavier than
iron even below the knee.
Since the number of muons was used to classify the showers, 
higher energies are needed with
SIBYLL to populate the same bin and consequently the
predictions for hadrons are higher, too.  However, when examining a quantity
which is independent of the muon number, SIBYLL performs much better as shown in
Fig. 12.
It is apparent that the results of the analysis depend strongly on the model.
In the case discussed here, the muon production in SIBYLL seems to underestimate
muon production, the
hadronic part, on the other hand, looks reasonable. 
This result is amazing, since muons
originate mainly from decaying charged pions, i.e. from the hadronic component.
The example illustrates the
power of multivariate measurements including the hadronic part of EAS.  Several
hadronic observables have been investigated and compared to various hadronic
interaction models \cite{HE1.2.27,drjoerg}.  Overall, CORSIKA calculations using
QGSJET and VENUS reproduce the experimental data best.

The KASCADE Collaboration presented four, still independent, analyses of the
composition from their data, comparing it to CORSIKA simulations with VENUS and
QGSJET.  One analysis uses the $N_e/N_\mu$ ratio as mass indicator
\cite{HE2.1.4}, in another the structure of the hadronic shower core is examined
\cite{HE1.2.27,drjoerg}, one employs the pattern of hits and reconstructed muons
in the central muon chambers \cite{OG6.1.35}, and in one multivariate analysis
techniques are used, combining several of the array observables available on an
event-by-event basis \cite{OG6.1.36}.  The results are shown in
Fig. 13. Though the results are still preliminary and some
inconsistencies have to be understood, all four analyses show qualitatively a
similar behaviour. There is a basically constant average mass for growing energy
up to about the knee region where a rise towards heavier masses sets in.

Another interesting result from the hadron calorimeter of the KASCADE experiment
was presented in Ref. \cite{HE1.2.28}. The longitudinal shower development
of single hadrons has been examined for hadrons well above the energies
accessible by man-made accelerators (see Fig. 14).  Thus,
cosmic-ray hadrons allow a test of the interaction codes used for detector
simulations in high-energy physics.  In general, GEANT simulations with the
FLUKA interaction model describe the data quite well up to energies of 10 TeV.
In addition, for energies $\ge$ 5 TeV, the shower attenuation is shown as measured
by Yakovlev et al. \cite{longfly} at Tien Shan. The authors see a flattening of
the attenuation with rising energy which they attribute to the existence of a
new {\em long flying component}.  The comparison with the KASCADE and the GEANT
shower curves, however, shows that there is reasonable agreement between them at
depths between 2 and 5 $\lambda_I$, i.e. in the range which was covered by the
Tien Shan detector. A small decrease of attenuation in this range of depth is
predicted by GEANT without any new physics, due to the shift of the shower
maximum with rising energy to larger depths.  Consequently, no indication of {\em new
physics} from hadronic shower curves can be established, at least up to 10 TeV,
by the KASCADE experiment.  In the near future the statistics of isolated 
high-energy hadrons in the KASCADE calorimeter will increase strongly, leading to
shower profiles for energies up to 50 TeV.
Further information on KASCADE results can be found in
Ref. \cite{highlight} in this volume.

\section{Emulsion Chamber Experiments}
A traditional method to investigate cosmic rays and high-energy hadronic
interactions is by means of emulsion chambers.  A stack of absorbers and sheets
of track detectors is exposed to cosmic radiation.  It is irradiated for
typically a year or two before the track detectors are processed.  Thus, the
definition of an event as an ensemble of tracks is purely based on geometry.
Emulsion chambers have a superb spatial resolution ($\approx\mu$m), however, the
x-ray films have a rather high threshold for photons in the TeV region 
and hadrons are not seen at all. Thus, the experiments suffer from  
the large fluctuations in the fraction of $\pi^0$s produced.  

Emulsion chambers are mostly operated at mountain altitudes or above the
atmosphere and aim for secondaries of a single interaction either in the
atmosphere above the chamber or in an absorber plate.

Typical setups have 2 storeys, the upper one for detection of photons from the
atmosphere and the lower one for detection of photons produced in the absorber
material of the upper layer.

Fig. 15 shows a schematic view of a cosmic-ray event as measured by an emulsion
setup and a typical pattern observable in one emulsion sheet.  The principle of
measurement with emulsion chambers was described recently in Ref. \cite{kempa}.

Energy determination of the photons is based on the optical density of the
film. The average darkness in a radius of 50 $\mu$m around a spot is related to
the energy deposit: The larger and the darker the spot, the higher is the energy
deposit. The spot size is determined by the contour line with optical density = 0.5. If
the area of the spot is $<$ 4 mm$^2$, the track is attributed to a single photon, if
the area is $>$ 4 mm$^2$, it is called a halo. An example of a halo event is shown in
Fig. 16. The optical densities in different layers
are combined to an estimate of the photon energy.  Since TeV photons are not
available at accelerators, the energy calibration cannot be verified experimentally.
Therefore, the response of X-ray films to $\gamma$ families has been
simulated recently using the CORSIKA and GEANT program packages \cite{HE1.2.29}.
The resulting relation between average optical density and photon energy is
shown in Fig. 17 for different zenith angles.  The shaded line
indicates the calibration function used by the PAMIR Collaboration for
experimental data.

The reconstruction of the energy $E_\gamma$ of the 
$\gamma$-family is a convolution of the
reconstruction quality for single photons and the energy and angular
distribution of the photons in an event.  On average the reconstruction using
the PAMIR calibration reproduces the ``true'' MC energy bias free, the
fluctuation extends up to 50\% as shown in Fig. 5 of Ref. \cite{HE1.2.29}.

\subsection{Aligned Events}
The PAMIR Collaboration observed a number of events in which the single particles of a
family or groups of particles form a straight line \cite{slavat}.  An
example of such an event is shown in upper part of 
Fig. 18 \cite{HE1.2.15}.

A strong alignment of the energy deposition was also found in the most
energetic event registered in emulsion chambers by the CONCORDE Collaboration
\cite{HE1.2.14}.  The emulsion chambers were exposed at flight altitude of the
CONCORDE aircraft which is 17 km corresponding to an overburden of 100 g/cm$^2$.
At those heights it is most likely that the secondaries observed in an event
originate from a single collision in the air above the plane.  The event is
sketched in the lower part of Fig. 18.  In total 1600 TeV have
been observed in this event giving a primary energy of $> 10^{17}$ eV. 90\% of
the energy is contained in 4 clusters that are aligned.  The 4 most energetic
individual photons (300, 105, 75, and 53 TeV) lie perfectly on a straight line.

Events of this type are expected to some extent. In $p$-$\bar{p}$ collisions at
CERN or Fermilab events with high $p_\perp$ jets are measured and well
understood in the framework of QCD. In the rest frame of one of the collision
partners such an event would produce automatically three coplanar jets entailing
three aligned groups of spots in a detector system.  More than 3 clusters may be
produced from initial states with high angular momentum.  String fragmentation
tends to produce particles in a plane, too.  Several other ideas have been put
forward to explain such events qualitatively.  Hence, the claim of an unusual
phenomenon requires a statistical analysis taking all these possible mechanisms
into account.

The PAMIR group has performed an analysis of the planarity of their families
using the two variables
$$
\lambda_N = \frac{\sum_{i\neq j\neq k} \cos 2\phi_{i,j}^k}{N(N-1)(N-2)}
\qquad \mbox{and} \qquad 
\alpha = \frac{\sum_{i\neq j} \cos 2\varepsilon_{i,j}}{N(N-1)} \quad .
$$
$\phi_{i,j}^k$ is the angle between the vector $(i,j)$ and the vector $(i,k)$
and $\varepsilon_{i,j}$ is the angle between the vector $(i,c)$ and the vector
$(j,c)$ where $c$ denotes the center of gravity of the gamma family. Both
variables are 0 for isotropic events and $\equiv 1$ for perfectly aligned events.

While Monte Carlo simulations using the programs MC0 and MJE reproduce the data
at $\sum E_\gamma < 400$ TeV, there is an excess of aligned events at a level of
2$\sigma$ for higher energies \cite{HE1.2.15,HE1.2.21}.  The excess of
alignment is present for events with three, four or five distinct particle groups.  By
combination of several distributions, a chance probability of such a result is
estimated to be $<10^{-5}$ \cite{slavat}.

In the case of the CONCORDE data, no such comparison has been performed yet.
The published event is the most 
energetic one and it is aligned. Thus, it is suggestive that
alignment may be a rather common feature at high energies.  

To further confirm the statement that alignment is due to new physics
with a threshold at about 400 TeV, it should be checked whether an
alternative interaction model, e.g. a high $p_\perp$ event generator from
collider physics, or some of the mechanisms proposed in the literature can
quantitatively explain the phenomenon. In addition, a thorough statistical
analysis of fluctuations and methodical or selection effects would be
necessary in order to prove whether the unusual events 
originate from a tail of a distribution or 
belong to a separate population. 

Due to the complicated handling and off-line processing of emulsions and X-ray
films and due to the very high-energy threshold of the films, an alternative
detection technique would be desirable to clarify the experimental
situation. Recent progress in detectors using scintillating fibers and
capillaries or in time-projection chambers using liquid active materials have
made a $\mu$m resolution feasible.

\subsection{Halo Events}
Altogether the PAMIR Collaboration found 39 events with halos for $\sum E_\gamma
> 500$ TeV (see Fig. 16).  These events were compared to MC
simulations of showers induced by single high-energy neutral pions (300-5000
TeV) and nuclear-electromagnetic cascades ($E_0 > 10$ PeV) \cite{HE1.2.19}.  The
Monte Carlo code used was MQ, which describes the main features of
$\gamma$-hadron families.  This code is based on the QGS model, but its details
are unpublished. The lateral distribution of darkness in the resulting spots for
data and simulations were compared and it is concluded that the large areas seen
in experimental data cannot be explained by simulations of high-energy $\pi^0$s
or shower cores.  However, it is crucial that the quoted program is able to
reproduce the response of the detector to a superposition of very many
low-energy subcascades as they occur in the core of a hadronic shower as well.

\subsection{Centauro Events}
Another exotic phenomenon reported repeatedly in the past are the Centauro events.
They were registered at Chacaltaya (Bolivia, 5200 m a.s.l.)   
with a two-storey
setup of emulsion chambers shown in Fig. 19. 
Both layers register only photons of TeV energies.  The upper layer is directly
sensitive to photons whereas spots in the lower layer are interpreted as
secondaries created from hadrons in the upper or the target layer 
and, thus, trace the number
of hadrons in the event.  For normal reactions in the atmosphere the expectation is
$$ 
\frac{N_{\gamma}/2}{N_{\rm h}} \approx 
\frac{N_{\pi^0}}{N_{\pi^++\pi^-}} \approx
\frac{1}{2} \quad .
$$  
Centauro events are those showing much fewer photons than expected ($Q_h = \sum
E_h / (\sum E_h + \sum E_\gamma) \approx 1$), Anti-Centauro events are those with
much more photons ($Q_h \approx 0$). The events are usually displayed in the
$N_h$-$Q_h$ plane as shown in Fig. 20. A new Centauro candidate
\cite{HE1.2.12} is plotted with a total energy of 57 TeV and $N_h = 13$, $Q_h
= 1.0$, i.e. with only hadrons and not a single photon above the threshold.
Fig. 20 also shows the distributions
of approximately the same number of simulated events as the data sample,
initiated by various primary nuclei. They exhibit a broad distribution and
scatter from $Q_h = 0$ to 1 due to fluctuations of the
$N_{\pi^0}/N_{\pi^\pm}$ ratio.  When assuming binomial fluctuations
in the number of charged and neutral pions one expects $8.7 \pm 1.7$ $\pi^\pm$
in an event with 13 pions in total.  The chance probability for no $\pi^0$ and
13 $\pi^\pm$ would then be $5\times 10^{-3}$.  From the plot containing about
4200 simulated events and none in the region near the measured event the chance
probability can be estimated to be $< 2\times 10^{-4}$, depending on the
interaction generator. Indeed, several authors claim to be able to reproduce
qualitatively Centauro-type events with suitable interaction models
\cite{attallah93,halzen}.
Whether the new event or the whole class of Centauros 
are unexplainable by known physics
can only be decided after a careful statistical study of the complete event set,
including the bias due to possible selection effects.  Such an analysis is very
difficult and has not yet been performed. As for the aligned events and halo
events, 
the data are
suggestive but the results are not convincing yet.

A dedicated search for anomalies in the fluctuations of the number of charged
pions in $p$-$\bar{p}$ collisions has been performed with the MINIMAX experiment
as mentioned in Sec. \ref{sec-minimax}.  No indications of unusual fluctuations
have been found so far.

\section{Conclusions}
The main progress over the past years in the field of high-energy interactions
and air shower physics consisted in the advent of better detector systems
allowing for multiparameter experiments with high-quality measurements of EAS
quantities and of powerful tools to investigate the shower development in the
atmosphere on the basis of elaborate theoretical models as well as the detailed
performance of the detectors used.  A standardization and improvement of those
tools to the best of our knowledge is a vital prerequisite for a consistent
comparison of results of different experiments.

We are convinced that a phase of improvement of the theoretical understanding, of
the methodical progress, and of the technical advancement will inevitably be
followed by a phase of interesting physics results.

\section*{Acknowledgments}
It is a pleasure to express my gratitude to the organizers of the 25$^{th}$ ICRC
for the lively and very well organized conference and, especially, for the invitation
to give a rapporteur talk and write this summary.

\noindent
I am indebted to numerous colleagues for patiently answering my questions,
giving many valuable comments and providing figures and transparencies during
the conference and the preparation of this article.

\noindent
In particular, I wish to thank J.N. Capdevielle, M.L. Cherry, B.R. Dawson,
L.W. Jones, J. Kempa, N.N. Kalmykov, G. Navarra, A. Ohsawa, E.H. Shibuya, and
S.A. Slavatinsky,
as well as my colleagues 
from the KASCADE Collaboration.

\noindent
My participation at the ICRC was supported by the Deutsche
Forschungsgemeinschaft.



\section*{Figures}

\noindent
Fig. 1 :
{\small left: pseudorapidity densities $\eta$ per wounded nucleon.
\quad  right: central pseudorapidity densities dN/d$\eta$ ($\eta \approx 3.25$) vs. 
total multiplicity $n_{prod}$ \protect\cite{HE1.1.6}. VENUS without re-interaction
reproduces the data better
than VENUS with re-interaction.}
\\[3mm]

\noindent
Fig. 2 :
{\small Proton structure function $F_2(x,Q^2)$ for low $Q^2$ and small $x$
as measured and predicted by various authors (see Ref. \protect\cite{h1pap} and
references therein)}
\\[3mm]

\noindent
Fig. 3 :
{\small Probability distributions for the $\pi^0$ production as obtained by 
binomial statistics (for 10 pions in total) 
and hadron production via DCC.}
\\[3mm]

\noindent
Fig. 4 :
{\small Layout of the FELIX detector.}
\\[3mm]

\noindent
Fig. 5 :
{\small Inelastic proton-air cross-sections as a function of energy 
according to experimental data
\protect\cite{mielke,gaisser_cross,yodh_cross,honda_cross,baltru_cross,HE1.2.1} 
and to the models in CORSIKA and MOCCA.}
\\[3mm]

\noindent
Fig. 6 :
{\small Comparison of the Fly's Eye and the Yakutsk data on
$X_{\rm max}$
with simulation of the QGSJET model for pure p and Fe (solid line) and for 
the $\sum$ composition (dashed line, see text) \protect\cite{kalm5}.}
\\[3mm]

\noindent
Fig. 7 :
{\small Comparison of the AKENO A1 data with simulations using MOCCA/SIBYLL for p
and Fe (left) and the resulting fraction of Fe (right) as a function of primary energy
\protect\cite{OG6.1.10}.}
\\[3mm]

\noindent
Fig. 8 :
{\small Is the composition changing or not? The answer depends on the
{\em yardstick} (i.e. Monte Carlo program) used for comparison.
Use the same yardstick to get {\em consistent} results,
use a well calibrated yardstick to obtain the {\em correct} result.}
\\[3mm]

\noindent
Fig. 9 :
{\small $N_\mu$ ($E_\mu > 1$ GeV) vs. $N_e$ from EAS-TOP for events
selected ``light'' or ``heavy'' on the basis of TeV muons in LVD, compared with
simulations for p and Fe nuclei \protect\cite{HE2.1.8}.}
\\[3mm]

\noindent
Fig. 10 :
{\small Flux of single hadrons as function of energy at 2200 and 110
m a.s.l. \protect\cite{HE1.2.23,mielke}.}
\\[3mm]

\noindent
Fig. 11 :
{\small Hadron energy spectra for measured air showers with $4.0 <
log_{10} N_\mu^{\rm tr} < 4.25$. left: comparison with VENUS predictions for
proton and iron primaries. \quad right: same for SIBYLL calculations
\protect\cite{HE1.2.27}.}
\\[3mm]

\noindent
Fig. 12 :
{\small Distribution of hadron energies normalized to the energy of
the most energetic hadron per event $E_{\rm h}/E_{\rm h}^{\rm max}$.  When
binning the events according to muon size (left) SIBYLL is not able to reproduce
the experimental data. When binning according to the hadronic energy sum (right)
there is fair agreement \cite{drjoerg}.}
\\[3mm]

\noindent
Fig. 13 :
{\small Four independent analyses of the composition by the KASCADE Collaboration
\protect\cite{HE2.1.4,HE1.2.27,drjoerg,OG6.1.35,OG6.1.36}.}
\\[3mm]

\noindent
Fig. 14 :
{\small Longitudinal shower development of hadrons in the KASCADE
hadron calorimeter as compared to GEANT/FLUKA simulations \protect\cite{HE1.2.28}.}
\\[3mm]

\noindent
Fig. 15 :
{\small left: Scheme of an air shower hitting the Pamir emulsion
chamber. Atmospheric photons are detected in the upper layer and hadrons produce
secondary photons which are detected in the lower layer (from Ref. \protect\cite{kempa}).
\quad right: Example of a $\gamma$-family in a X-ray film (from S.A. Slavatinsky).}
\\[3mm]

\noindent
Fig. 16 :
{\small Example of an event with a large halo (from S.A. 
Slavatinsky). It extends over
several cm$^2$ corresponding to an energy of 50 PeV.  The energy of the primary
producing this event is estimated to be about 1000 PeV.}
\\[3mm]

\noindent
Fig. 17 :
{\small Average optical density vs photon energy as obtained from
GEANT simulati ons for different zenith angles. The errors of the points are
typically 0.2 \protect\cite{HE1.2.29}.  The shaded line indicates the
calibration used by the PAMIR Collaboration.}
\\[3mm]

\noindent
Fig. 18 :
{\small top: Example of a PAMIR event with aligned particle groups
\cite{HE1.2.15}. \quad bottom: The most energetic CONCORDE event
\cite{HE1.2.14}. The circles indicate groups of individual particles.}
\\[3mm]

\noindent
Fig. 19 :
{\small The emulsion chamber setup at Chacaltaya.} 
\\[3mm]

\noindent
Fig. 20 :
{\small Distribution of simulated events in the $N_h$-$Q_h$
plane. The plot contains 4200 events of different primaries 
with $30 \le \sum E_\gamma < 100$ TeV.  The
new Centauro candidate is indicated with the large black dot.}

\end{document}